\documentclass[sigplan, nonacm]{acmart}

\usepackage{microtype}
\microtypesetup{expansion=false,protrusion=false}

\usepackage{hyperref}

\usepackage{hyphenat}

\renewcommand\footnotetextcopyrightpermission[1]{}
\makeatletter
\@ACM@balancefalse
\makeatother

\AtBeginDocument{%
	}

\acmConference[Agent Skills'26]{ACM CAIS Workshop on Design, Evaluation, and Optimization of Procedural Knowledge for LLM Agents}{May 26, 2026}{San Jose, CA, USA}

\begin{document}
	
	\title{Toward User Comprehension Supports for LLM Agent Skill Specifications}
	\subtitle{Rare User-Facing Demonstrations Across 878 Cybersecurity Skills}
	
	\author{Zikai Alex Wen}
	\orcid{0000-0001-9163-7450}
	\affiliation{%
		\institution{University of Washington, Tacoma School of Engineering \& Technology}
		\country{Tacoma, Washington, USA}
	}
	\email{zkwen@uw.edu}
	
	\begin{abstract}
		Users often interpret and select agent skills through their \texttt{SKILL.md} specifications. To protect users, existing audits mainly focus on malicious or unsafe skills. We study the complementary question of whether specifications help users form bounded expectations about what a skill consumes, produces, and covers. Across 878 cybersecurity skills, we used rule-based coding to measure textual cues for four comprehension anchors, namely operational basis, output contract, boundary disclosure, and example capability demonstration. Cues for operational basis were common, but only 19.0\% of specifications exhibited cues for an example task, sample, or expected outcome, and only 2.3\% exhibited cues for all four anchors. We further examined a small DNS/C2 telemetry subset (n$=$6) to illustrate why missing examples may matter. Examples appeared to make first local checks easier to construct, while no-example skills typically required helper code inspection to recover command arguments or output fields. We argue that agent-skill evaluation should treat specifications as user-facing capability disclosures, not merely as containers for executable instructions.
	\end{abstract}

	
	\maketitle
	
	\section{Introduction}
	
	Agent skills have emerged as a distribution format for procedural knowledge that LLM agents can load on demand. Public repositories now distribute thousands of YAML and Markdown skill packages for compatible agents~\cite{lardinois2025Agent}. Cybersecurity is one of the popular categories, covering secure code review, CWE triage, DNS tunneling detection, and threat-intelligence extraction. The stakes are high because users may install these skills to perform tasks they cannot independently verify.
	
	In this setting, the user-facing \texttt{SKILL.md} file does double duty. It is an instruction file that tells an agent how to perform a task, but it is also the document that people read when deciding whether a skill fits their task and risk tolerance. A skill can describe how to review source code, analyze DNS logs, or produce a structured security report. If the specification does not convey what the skill is for, what evidence it expects, and what result the agent may produce, users can run skills they should not, or trust outputs that exceed the skill's actual coverage.
	
	Recent LLM agent skill audits~\cite{heady2026HeadyZhang, pors2026Skillaudit, safo-lab2026DynAuditClaw} have examined \texttt{SKILL.md} files for adversarial behavior, including prompt injection, unsafe tool use, and malicious behavior, with the goal of protecting users from unsafe or malicious skills. Nonetheless, an audit for malicious behavior is not the same as a comprehension aid. 
	
	Showing that a skill is non-malicious does not ensure that users can run it with appropriate expectations. We drew on description-to-permission fidelity research, which compared user-facing app descriptions against technical access~\cite{panditaWHYPER,qu2014AutoCog,shezan2020TKPERM}, as a methodological analogue. For agent skills, the relevant comparison is not between descriptions and permissions, but between specification text and operational capability.
	
	We investigated a concrete precondition for user comprehension. Can a user form a usable expectation about the skill and construct a first local check from the specification itself? This precondition aligns with prior findings on usable security disclosures, which succeed only when users can attend to, understand, and act on them~\cite{felt2012Android,schaub2015Design,kelley2009Nutrition}. A skill that supports comprehension does not merely state ``detect DNS tunneling''. It may name the expected input artifact, state the output shape, disclose relevant limits, and provide a small example, fixture, test case, or expected output that makes the behavior concrete. A less supportive skill may still work, but the user may have to inspect helper code or infer hidden assumptions. 
	
	Our contribution is threefold. Firstly, we characterized four user-facing comprehension supports in agent skill specifications and quantified their prevalence across 878 cybersecurity \texttt{SKILL.md} files. Secondly, we examined a local DNS/C2 telemetry subset and observed that example demonstrations made first local checks more legible from the specification. Thirdly, we used these findings to identify emerging questions about skill-specification design and user comprehension supports.
	
	\section{Related Work}
	
	Our work builds on three lines of prior research. Existing skill-safety audits targeted the same artifact we study (i.e., \texttt{SKILL.md}), but focused on adversarial behavior, such as hidden instructions, unsafe tool use, or attempts to leak data. User-facing comprehension was largely outside their scope. In contrast, research on security and privacy disclosures shows that user comprehension is critical and should be treated as an essential evaluation criterion as well. Building on this insight, we adapted the logic of description-to-permission fidelity from app ecosystems to ask whether specifications expose the operational anchors users need to evaluate a skill. 
	
	\subsection{Skill-Safety Audits}
	
	Recent projects~\cite{heady2026HeadyZhang, pors2026Skillaudit, safo-lab2026DynAuditClaw} have emerged to audit agent skills for security risks in their specifications and helper code. For example, Agent Audit~\cite{heady2026HeadyZhang} and skill-audit~\cite{pors2026Skillaudit} statically analyzed \texttt{SKILL.md} files for adversarial behavior. In this context, adversarial behavior refers to specification content that tries to steer the agent away from the user's intent or security policy, for example by embedding prompt-injection instructions, requesting dangerous shell or network actions, concealing side effects, or exfiltrating secrets. DynAuditClaw~\cite{safo-lab2026DynAuditClaw} extended this approach to dynamic execution by cloning the agent installation into an isolated container and judging adversarial scenarios against a three-axis taxonomy. Our study complements this body of work by examining whether a non-adversarial skill specification provides users with sufficient information to construct an initial capability check.
	
	\subsection{Disclosure Comprehension}
	
	Research on security and privacy disclosures shows that notices are most useful when they support user attention, comprehension, and decision-making. Felt et al.~\cite{felt2012Android} found that Android permission prompts received little attention and were poorly understood by many users. Schaub et al.~\cite{schaub2015Design} systematized the design space for privacy notices and emphasized usable, useful notice for different audiences and contexts. Labels can improve access to information, but confusing terminology and information structure can impair comprehension~\cite{kelley2009Nutrition,zhang_exploring_2024,keswani_user_2025}. In this work, we treated \texttt{SKILL.md} as an analogous notice artifact for agent capabilities.
	
	\subsection{Description-to-Permission Fidelity}
	
	WHYPER~\cite{panditaWHYPER} introduced an NLP pipeline for identifying whether Android app descriptions explained the need for sensitive permissions, framing the task as a way to bridge the semantic gap between user expectations and application functionality. AutoCog~\cite{qu2014AutoCog} generalized this into ``description-to-permission fidelity'' and measured low-fidelity descriptions across 45{,}811 Android applications. TKPERM~\cite{shezan2020TKPERM} later transferred permission knowledge across Android, Chrome extensions, and IFTTT, showing that vague descriptions and sensitive-access decisions were a cross-platform problem. More recent work has continued to identify mobile permission-disclosure discrepancies and limits in user awareness~\cite{donku2025Discrepancies}. We adopted the core idea of comparing user-facing descriptions with technical behavior, while adapting the measurement target. 
	
	\section{Methods}
	
	\subsection{Corpus Coding of Comprehension Supports} 
	
	We analyzed 878 \texttt{SKILL.md} files across five GitHub sources, including two institutional sources (Transilience~\cite{transilience2026Community} and Trail of Bits~\cite{trailofbits2026Skills}) and three community sources (mukul975~\cite{jangra2026Anthropic}, alirezarezvani~\cite{rezvani2026Claude}, and Eyadkelleh~\cite{kelleh2026Awesome})\footnote{Source URLs, coding scripts, and output data are available at \url{https://github.com/zikaiwen/cyber-skill-comprehension}}. We treated each \texttt{SKILL.md} as a capability disclosure, a document that should let a reader form a testable expectation about what the skill does. This framing draws on disclosure-comprehension work~\cite{felt2012Android,schaub2015Design,kelley2009Nutrition}, which argues that notices are useful only when users can understand them and act on them. It also draws on description-to-permission fidelity work~\cite{panditaWHYPER,qu2014AutoCog,shezan2020TKPERM,watanabe_understanding_2015}, which measures whether descriptions explain the technical behavior that users are asked to trust. We adapted these ideas from permissions to agent skills. Our question is whether the specification gives enough anchors for a reader to understand the skill's capability and construct a first check.
	
	We operationalized this expectation as four anchors. \emph{Operational basis} asks whether the specification tells the reader what evidence, inputs, tools, or prerequisites the skill relies on. \emph{Output contract} asks whether the reader can anticipate what kind of result the skill should produce and judge whether an agent's answer is on-task. \emph{Boundary disclosure} asks whether the specification states where the capability does or does not apply. \emph{Example capability demonstration} asks whether the specification gives a concrete example, sample output, or test case that could seed a first local check.
	
	Because this was a corpus-scale study, we did not code these anchors as deep semantic judgments about whether a skill truly works. Instead, we measured observable textual cues that would plausibly expose each comprehension anchor to a reader. Table~\ref{tab:coding-rules} summarizes the cue families used to code the anchors. We implemented the coding as a deterministic, rule-based scan over each full \texttt{SKILL.md}. All text patterns were case-insensitive, and heading patterns also used multiline matching.
	
	\begin{table}[t]
		\centering
		\footnotesize
		\caption{Coding rule families for corpus analysis.}
		\label{tab:coding-rules}
		\begin{tabular}{p{0.30\linewidth}p{0.62\linewidth}}
			\toprule
			\textbf{Variable} & \textbf{Cue families} \\
			\midrule
			Operational basis & Input-oriented headings such as \texttt{Prerequisites}, \texttt{Inputs}, \texttt{Data Sources}, \texttt{Evidence}, \texttt{Artifacts}, \texttt{Tools}, and prose where verbs such as ``requires'', ``expects'', ``accepts'', or ``consumes'' introduce files, logs, URLs, endpoints, or other input artifacts. \\
			Output contract & Output-oriented headings and phrases naming report structures, result schemas, finding fields, or concrete emitted artifacts such as JSON, CSV, SARIF, Markdown, PDF, YAML, STIX, Sigma, YARA, reports, summaries, dashboards, tables, artifacts, or findings. \\
			Boundary disclosure & Explicit limits and scope cues, including \texttt{Limitation}, \texttt{Scope}, \texttt{Coverage}, \texttt{out of scope}, \texttt{does not}, \texttt{not supported}, \texttt{not intended for}, false-positive/negative language, caveats, exclusions, and assumptions. \\
			Example capability demonstration & Example-oriented headings and phrases, including \texttt{Example}, \texttt{Usage}, \texttt{Sample Input}, \texttt{Sample Output}, \texttt{Quickstart}, \texttt{Test Case}, \texttt{Demo}, \texttt{Walkthrough}, \texttt{expected output}, \texttt{fixture}, \texttt{synthetic example}, and \texttt{try this}. \\
			\bottomrule
		\end{tabular}
	\end{table}
	
	\subsection{Local DNS/C2 First-Check Subset} 
	
	After the corpus coding, we examined the example capability demonstration anchor more closely in a narrow local DNS/C2 subset. The goal was to see whether examples made a first local check easier to construct from \texttt{SKILL.md} before reading helper code. Helper code refers to executable files shipped with a skill package, such as local scripts that parse telemetry and produce report artifacts invoked by \texttt{SKILL.md}. We included skills whose title or description names DNS tunneling, DNS exfiltration, command-and-control beaconing, or ransomware network indicators; whose package includes helper code that accepts Zeek, DNS, NetFlow, or TSV/CSV telemetry files; and whose local check does not require cloud credentials, external services, or unavailable third-party Python packages. This selection yielded six skills. For this subset, we treated an example as a concrete sample input, sample output, or run result. Only one skill provided such an example, although some named expected output categories or detection workflows.
	
	We created small synthetic fixtures, including a Zeek \texttt{dns.log} containing high-entropy DNS queries, a Zeek \texttt{conn.log} with regular beacon-like intervals, and plain TSV variants for helper code parsers that expected header-first tabular files. For each skill, we posed a direct local agent task, such as using the skill to analyze the fixture for DNS tunneling. We recorded whether \texttt{SKILL.md} exposed sufficient information to identify the input artifact, expected output shape, and first checkable behavior before inspecting helper code. We then used the helper code only as the skill package's backend execution mechanism.
	
	\section{Preliminary Findings}
	
	We present preliminary findings on whether \texttt{SKILL.md} files support user comprehension. The corpus analysis measures the distribution of four comprehension anchors across 878 cybersecurity skills, and the DNS/C2 subset suggests how missing examples can move first-check construction from specification reading to helper code inspection.
	
	\subsection{Uneven Comprehension Anchors}
	
	The corpus results show that skill specifications often gave readers some basis for forming expectations, but not a complete path from input to bounded, checkable behavior. Operational basis was common. In total, 92.1\% (809/878) named inputs, prerequisites, evidence, data sources, tools, or validation in a user-facing section. Output contracts and boundary disclosures were less consistent. Only 63.0\% (553/878) specified a named output or report artifact, and 51.4\% (451/878) disclosed coverage or limitations. Example demonstrations were much rarer, appearing in just 19.0\% (167/878) of skills. Table~\ref{tab:summary} reports these percentages of anchor presentation.
	
	The anchors were also rarely bundled. Only 2.3\% (20/878) provided all four anchors together. This gap suggests that authors did not typically produce comprehension support as a coordinated package. Many skills named the materials, evidence, or tools involved, but far fewer carried that description through to an expected result, a boundary statement, and a concrete first check.
	
	\begin{table}[t]
		\centering
		\footnotesize
		\caption{Corpus-coded comprehension anchors.}
		\label{tab:summary}
		\begin{tabular}{lr}
			\toprule
			\textbf{Measure} & \textbf{Share} \\
			\midrule
			Operational basis disclosed & 92.1\% \\
			Named output contract present & 63.0\% \\
			Boundary disclosure present & 51.4\% \\
			Example capability demonstration present & 19.0\% \\
			All four anchors present & 2.3\% \\
			\bottomrule
		\end{tabular}
	\end{table}
	
	\subsection{Examples Made First Checks More Legible}
	\label{sec:spotcheck}
	
	The local DNS/C2 subset examined the mechanism introduced in Methods by asking whether an example anchor makes the input, output, and first check visible from the specification itself. This analysis was not a direct user-comprehension study and was not a detector benchmark. Instead, it asked what a reader could recover from \texttt{SKILL.md} before inspecting helper code, including the expected input artifact, the shape of the output or check, and a plausible first local task.
	
	Only \texttt{dns-exfil-zeek} provided a concrete package-level example: an expected JSON output with summary fields, flagged domains, and evidence indicators. The five no-example skills were usable, but their specifications left varying amounts of implementation recovery to helper code, ranging from unspecified JSON fields to undocumented argument names, missing minimal log examples, and implicit parser expectations (listed in Table~\ref{tab:local-comprehension}).
	
	\begin{table}[t]
		\centering
		\footnotesize
		\caption{Information that still required code inspection before constructing a first local check in the DNS/C2 subset.}
		\label{tab:local-comprehension}
		\begin{tabular}{@{}p{0.32\linewidth}p{0.14\linewidth}p{0.42\linewidth}@{}}
			\toprule
			\textbf{Skill Name} & \textbf{Example} & \textbf{What remained unclear?} \\
			\midrule
			\texttt{dns-exfil-zeek} & Y & No substantial recovery needed \\
			\texttt{ransomware-net} & N & Exact JSON fields not shown \\
			\texttt{dns-query-analysis} & N & Runnable command and output fields \\
			\texttt{beaconing-freq.} & N & Minimal connection log and expected beacon report \\
			\texttt{data-exfil-ind.} & N & Header-first TSV parser expectation \\
			\texttt{dns-tunnel-zeek} & N & Runnable command and emitted output fields \\
			\bottomrule
		\end{tabular}
	\end{table}
	
	This subset does not show that the example-bearing skill was more correct. It suggests that a concrete sample output made a first local check easier to construct from the specification. Weaker supports, such as output-category bullets, could still help readers anticipate the result, but they did not provide the same checkable anchor as a concrete example.
	
	\section{Discussion and Emerging Questions}
	\label{sec:agenda}
	
	Our findings make examples a useful site for further study. In the DNS/C2 subset, examples helped make inputs and outputs concrete enough to construct a first local check. But examples are not equivalent to unit tests, and they do not by themselves establish correctness, coverage, or boundary behavior. This raises three potential research questions about when skill examples help, when they mislead, and how they relate to broader trust signals in a skill ecosystem.
	
	\textbf{Q1. When do examples improve comprehension?} Our findings suggest that examples can reduce the work of constructing a first local check, but they do not establish that examples are always necessary. Some skills may be understandable through a clear operational basis, output contract, or boundary disclosure alone. Building on prior work on permission comprehension and privacy-label terminology~\cite{felt2012Android,keswani_user_2025}, future user studies could vary examples against other specification supports and measure attention, comprehension, risk perception, and expectation accuracy.
	
	\textbf{Q2. When do examples mislead?} Examples, when present, are also skill capability claims. A selected fixture or sample output can clarify intended behavior, but it can also exaggerate coverage, omit boundary cases, or show an output that the helper code does not reliably produce. Future work may examine when examples help users form accurate expectations and when they instead function as persuasive but incomplete evidence. Prior permission-correlation systems demonstrated the value of matching natural-language descriptions against technical access~\cite{panditaWHYPER,qu2014AutoCog,shezan2020TKPERM,watanabe_understanding_2015}. Agent skills invite a related check between examples, specifications, and executable behavior. Tools could support this by comparing examples against helper code, output schemas, and synthetic counterexamples. This direction complements security-focused skill audits~\cite{heady2026HeadyZhang,pors2026Skillaudit,safo-lab2026DynAuditClaw} by surfacing non-adversarial skills that are still difficult for users to evaluate.
	
	\textbf{Q3. Where should comprehension support live?} The need for examples may depend on the surrounding ecosystem. App stores and browser-extension markets often let users rely on ratings, reviews, install counts, publisher reputation, screenshots, permissions, and platform moderation rather than examples for every capability~\cite{harman_app_2012,olsson_fakex_2024}. Agent-skill ecosystems may develop similar signals. However, these signals may not reveal the task interface of a specific skill, such as telemetry formats, output fields, parser assumptions, or boundary conditions. They may also become an attack surface when manipulated ratings, reviews, or reputation cues encourage users to skip capability-specific inspection. Future work should examine how reputation and review mechanisms can be combined with specification-internal supports to nudge users from high-level trust cues toward concrete evidence before use.
	
	\section{Limitations}
	
	The findings of this study are subject to at least four limitations. First, our coding measures textual cues that plausibly support comprehension, not user comprehension itself. Thus, a controlled user study is needed to test whether the anchors translate into accurate expectations. Second, the rule-based coding was not validated against manual annotation, so reported percentages should be read as approximate prevalence estimates. Third, the DNS/C2 subset is illustrative and was not designed to support general claims about example demonstrations. Fourth, our corpus is cybersecurity-focused, and comprehension patterns in other high-stakes skill domains may differ.
	
	\section{Conclusion}
	
	This paper reframes agent-skill evaluation around whether \texttt{SKILL.md} specifications help users form bounded expectations before running a skill or interpreting its output. Across 878 cybersecurity skills, operational basis was common, but output contracts, boundary disclosures, and example demonstrations were less consistently provided; only 2.3\% included all four anchors. The DNS/C2 subset suggests that examples can make a first local check easier to construct by making inputs and outputs concrete, while also underscoring that examples are not tests and should not be treated as proof of coverage or correctness. We therefore recommend studying and auditing specification supports, including named outputs, boundary disclosures, and carefully scoped examples, as part of trustworthy agent-skill ecosystems.

	\bibliographystyle{ACM-Reference-Format}
	\bibliography{reference}
	
\end{document}